\RequirePackage{fix-cm}
\documentclass[twocolumn]{svjour3}          
\smartqed  
\usepackage{graphicx}
\usepackage[pdfborder={0 0 0},colorlinks=true,linkcolor=blue,citecolor=blue]{hyperref}

\def\beq{\begin{equation}}
\def\eeq{\end{equation}}
\def\bea{\begin{eqnarray}}
\def\eea{\end{eqnarray}}

\newcommand{\STO}{SrTiO$_3$~}

\begin{document}

\title{Effect of the additional Se layer on the electronic structure of iron-based superconductor FeSe/SrTiO$_3$ \thanks{This work was supported in part by ``BASIS'' Foundation for Development of Theoretical Physics and Mathematics.}
}
\subtitle{}


\author{L.V. Tikhonova \and M.M. Korshunov}


\institute{L.V. Tikhonova and M.M. Korshunov \at
               Siberian Federal University, Svobodny Prospect 79, 660041 Krasnoyarsk, Russia \\
           \and
           M.M. Korshunov \at
              Kirensky Institute of Physics, Federal Research Center KSC SB RAS, Akademgorodok, 660036, Krasnoyarsk, Russia \\
              \email{mkor@iph.krasn.ru}
}

\date{Received: date / Accepted: date}

\maketitle

\begin{abstract}
We use density functional theory to study the structure and the band structure of the monolayer FeSe deposited on the \STO substrate with the additional layer of Se between them. Top of the \STO is formed by the double TiO layer with and without oxygen vacancies. Several structures with different arrangements of the additional Se atoms above the double TiO layer is considered. Equilibrium structures were found and the band structures for them were obtained. Near the $\Gamma=(0,0,0)$ point of the Brillouin zone, the hole Fermi surface pockets persist and, additionally, an electron pocket appears. Thus neither the presence of the additional Se layer nor the oxygen vacancies in the double TiO layer leads to the sinking of hole bands below the Fermi level near the $\Gamma$ point. Necessity to include the strong electronic correlations into account is discussed.
\keywords{Fe-based superconductors \and Monolayer FeSe \and SrTiO$_3$ \and density functional theory}
\end{abstract}

\section{Introduction}
\label{intro}
Iron-based materials represents a class of high-$T_c$ superconductors that is based on the conducting layer of Fe surrounded by As or Se in pnictides and chalcogenides, respectively \cite{y_kamihara_08,SadovskiiReview2008,IzyumovReview2008,IvanovskiiReview2008,JohnstonReview,PaglioneReview,LumsdenReview,StewartReview,Inosov2016}. Each of the constituents can be replaced by some other element thus providing a doping and/or crystal lattice distortion. This results in the rich phase diagram that includes, among others, antiferromagnetic and superconducting phases. Proximity of the magnetic phase implies the role of spin fluctuations in the formation of Cooper pairs \cite{HirschfeldKorshunov2011}. Stemming from the ideas of Berk and Schrieffer \cite{BerkSchrieffer}, the theory of spin fluctuation-mediated pairing provides consistent explanation of several important features of superconductivity in Fe-based materials \cite{HirschfeldKorshunov2011,ChubukovReview2012,Korshunov2014eng}. Observation of the spin resonance peak in inelastic neutron scattering \cite{KorshunovEreminResonance2008,Maier2008,Maier2009,KorshunovPRB2016,Korshunov2019} and the quasiparticle interference pattern \cite{Akbari2010,Wang2010,AkbariQPI2010,Hirschfeld2015} in scanning tunneling spectroscopy confirms sign-changing $s_\pm$ gap predicted by the spin fluctuation theory of superconductivity \cite{Mazin2008,Chubukov2008,Kuroki2008,Graser2009,MaitiKorshunovPRL2011,MaitiKorshunovPRB2011,MaitiKorshunov2012,Classen2017}. Other approaches to the theory includes pairing due to the orbital fluctuations supported by phonons \cite{Kontani} or the Aslamasov-Larkin vertex corrections \cite{Onari2012}.

Discovery of superconductivity in FeSe monolayers with $T_c$ as high as 100~K presents another mystery in the physics of iron-based materials \cite{FeSeTc,Zhang2015,GeFeSe100K,ZhaoFeSeLiOHFeSe,Sadovskii2016}.
Different mechanisms of superconductive pairing in monolayer FeSe were suggested including specific realization of the electron-phonon interaction \cite{Lee2014,Coh2015,Gorkov2016,Kulic2017}, nematic fluctuations \cite{Kang2016,FernandesReview2017,Kang2018}, orbital fluctuations \cite{Yamakawa2017}, spin fluctuations \cite{Kreisel2017}, and spin fluctuations involving incipient bands \cite{Chen2015}. Recent advances in scanning tunneling spectroscopy of iron chalcogenides support sign-changing gap \cite{Du2017,Liu2019,Jandke2019}.

Any pairing theory is tightly connected to the underlaying band structure and the Fermi surface. Thus, knowledge of the band structure details becomes extremely important. Angle-resolved photoemission spectroscopy (ARPES) shows that there is only one kind of Fermi surface sheets -- electron pockets around the corners of the Brillouin zone ($M$ points), while there is no hole pockets around the center ($\Gamma$ point) \cite{FeSeARPES}. Moreover, with increasing number of FeSe layers hole pockets around the $\Gamma$ point become visible in ARPES and at the same time the superconductivity disappears \cite{TanFeSeARPES}. Surprisingly, despite a successful qualitative and even semi-quantitative description of the bulk Fe-pnctides and Fe-chalcogenides electronic structure within the modern band structure calculation schemes, they fail to reproduce such a topology of the Fermi surface. For example, existing density functional studies do not reveal significant differences between the monolayer and double unit cell layer of FeSe \cite{LiuPRB2012}. There are at least two reasons why this may happen. First one is related to the global problem of the density functional theory (DFT) in describing the systems with strong electronic correlations (SEC). If this is the case, i.e. the monolayer FeSe represents a class of systems with SEC, then some specific approach is required to describe its electronic structure and, moreover, theory of superconductivity should be formulated taking SEC into account. There is, however, another possibility of why there is a disagreement between the results of DFT and the experimentally observed Fermi surface. It is related to the atomic structure of the FeSe/\STO interface. The interface seems to be more complicated then can be naively expected because the preparation of the monolayer FeSe is quite cumbersome. If FeSe deposited on graphene or \STO it doesn't become neither metal nor superconductor. However, if it is deposited on Se-enriched \STO and then annealed, the superconductivity with high $T_c$ appears \cite{FeSeTc,GeFeSe100K}. The process of annealing and Se bombardment naturally modifies the surface layer of \STO. There are few experimental studies of the interface \cite{Li2DMat2016,Zou2016}, which revealed the presence of the TiO$_2$ double layers on top of \STO. Both this and the possible oxygen vacancies due to the annealing were incorporated in the first principles calculations \cite{Bang2013,Xu2017}. They had some success in explaining the disappearance of the hole pocket around the $\Gamma$ point. Recent experimental study \cite{Zhao2018} utilized the high-angle annular dark-field scanning transmission electron microscopy (HAADF-STEM) images and revealed an additional Se layer located between FeSe monolayer and TiO$_x$-terminated \STO. This new fact may be important for the FeSe band structure and have to be taken into account.

Here we use DFT calculations to study the influence of the additional Se layer on FeSe/(2TiO)\STO electronic structure. Several possible positions of additional Se atoms and of FeSe monolayer in the unit cell of $\mathrm{FeSe}/\mathrm{Se}/(2\mathrm{TiO}_x)\mathrm{SrTiO}_3$ heterostructure were modeled to search for favorable configurations and the electronic structure of these configurations were investigated.

The paper is organized as follows. Section~\ref{sec:caclulation} presents details of the computational procedure, Section~\ref{sec:results} contains results and their discussion, and Section~\ref{sec:conclusion} is the conclusion.

\begin{figure}
 \includegraphics[width=0.99\columnwidth]{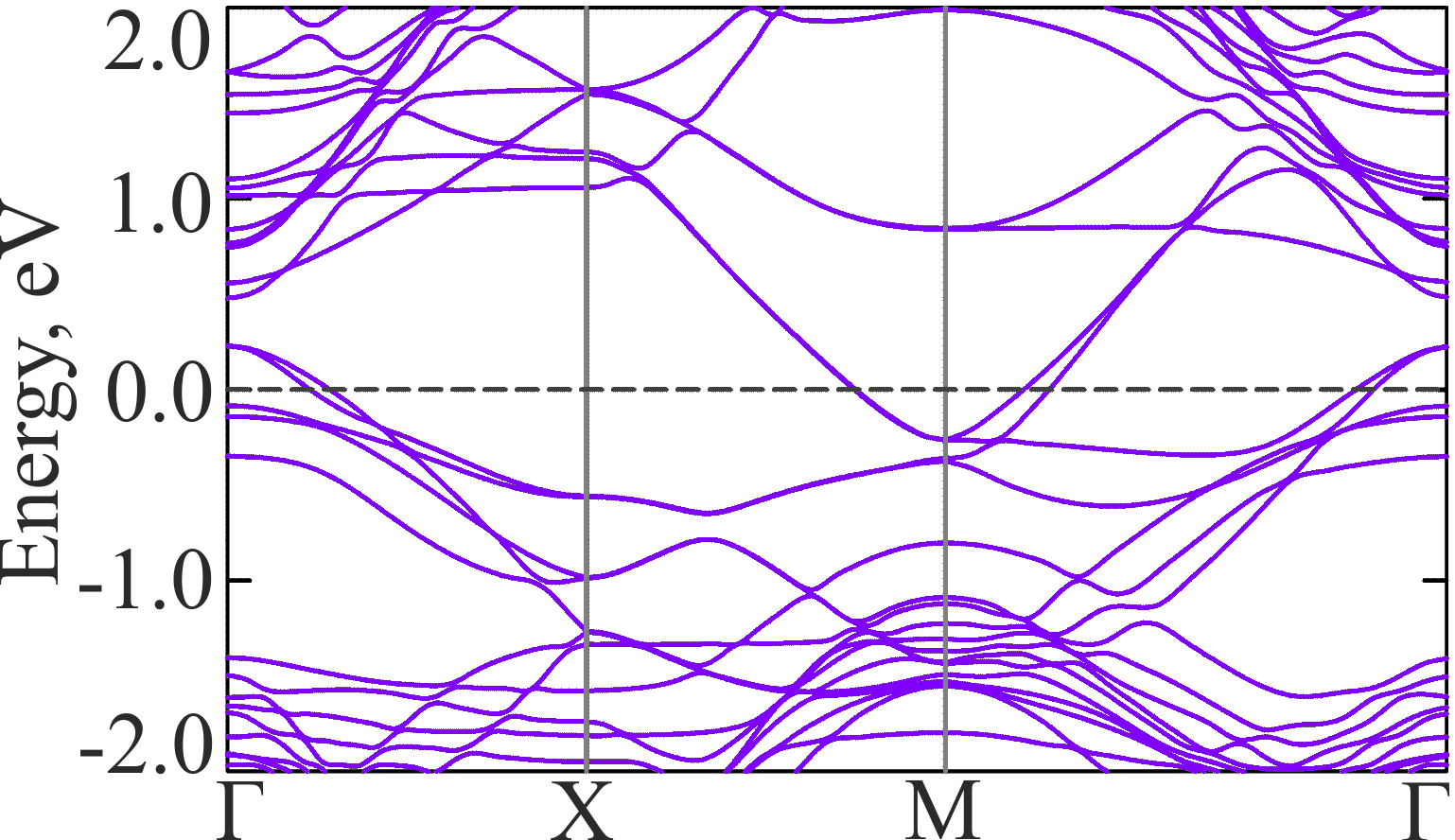}
 \caption{Band structure of the FeSe/\STO with the double TiO layer without vacancies. Fermi level is set to zero.}
\label{fig:ref}
\end{figure}

\section{Computational details}
\label{sec:caclulation}
DFT \cite{p_hohenberg_64,wkohn65} calculations were carried out using Open source package for Material eXplorer software package (OpenMX, \url{http://www.openmx-square.org})
based on a linear combination of pseudoatomic orbital (PAO) method \cite{Ozaki2004} and norm-conserving pseudopotentials \cite{Bachelet1982,Kleinman1982,Blochl1990,Troullier1991,Morrison1993}. The cutoff energy value was equal to 150~Ry. The PAO basis set with $s2p2d2f1$ for Fe and Ti, $s2p2d2$ for Se and O, and $s2p1$ for Sr were found to be sufficient to describe the structures. Cutoff radii of 10.0~a.u. for Sr, 7.0~a.u. for Ti and Se, 6.0~a.u. for Fe, and 5.0~a.u. for O were used. Exchange-correlation effects were taken into account in the generalized gradient approximation (GGA) by the Perdew-Burke-Ernzerhof (PBE) functional \cite{jperdew96}. The structures were relaxed until the forces acting on atoms and the translation vectors became less then $1 \cdot 10^{-4}$ Hartree/Borh. The convergence condition for the energy was $1 \cdot 10^{-6}$ Hartree. To simulate the two-dimensional (2D) structures using Periodic Boundary Conditions (PBC), the periodic replicas were separated by a vacuum spacing of at least 20~\AA along the $c$ axis. The first Brillouin zone (BZ) was sampled with $12 \times 12 \times 1$ $k$-grid. Band structure calculations were performed along the high symmetry directions in BZ: $\Gamma(0,0,0) - X(1/2,0,0) - M(1/2,1/2,0) - \Gamma(0,0,0)$. To avoid the edge effects on the atomic and electronic structures, the \STO supercell $1 \times 1 \times 4$ was used in calculations. The Visualization for Electronic and Structural Analysis (VESTA) software \cite{Momma2011} was used to visualize the atomic structures.

\section{Results and discussion}
\label{sec:results}
In Fig.~\ref{fig:ref}, we show the band structure of the reference system, i.e. the FeSe/(2TiO)\STO heterostructure without vacancies and without the additional Se layer. Notice the bands slightly above the Fermi level ($E_F$) at the $\Gamma$ point, which have downward dispersion thus forming the hole Fermi surface pockets in the center of the BZ.

\begin{figure}
\begin{center}
 \includegraphics[width=0.8\columnwidth]{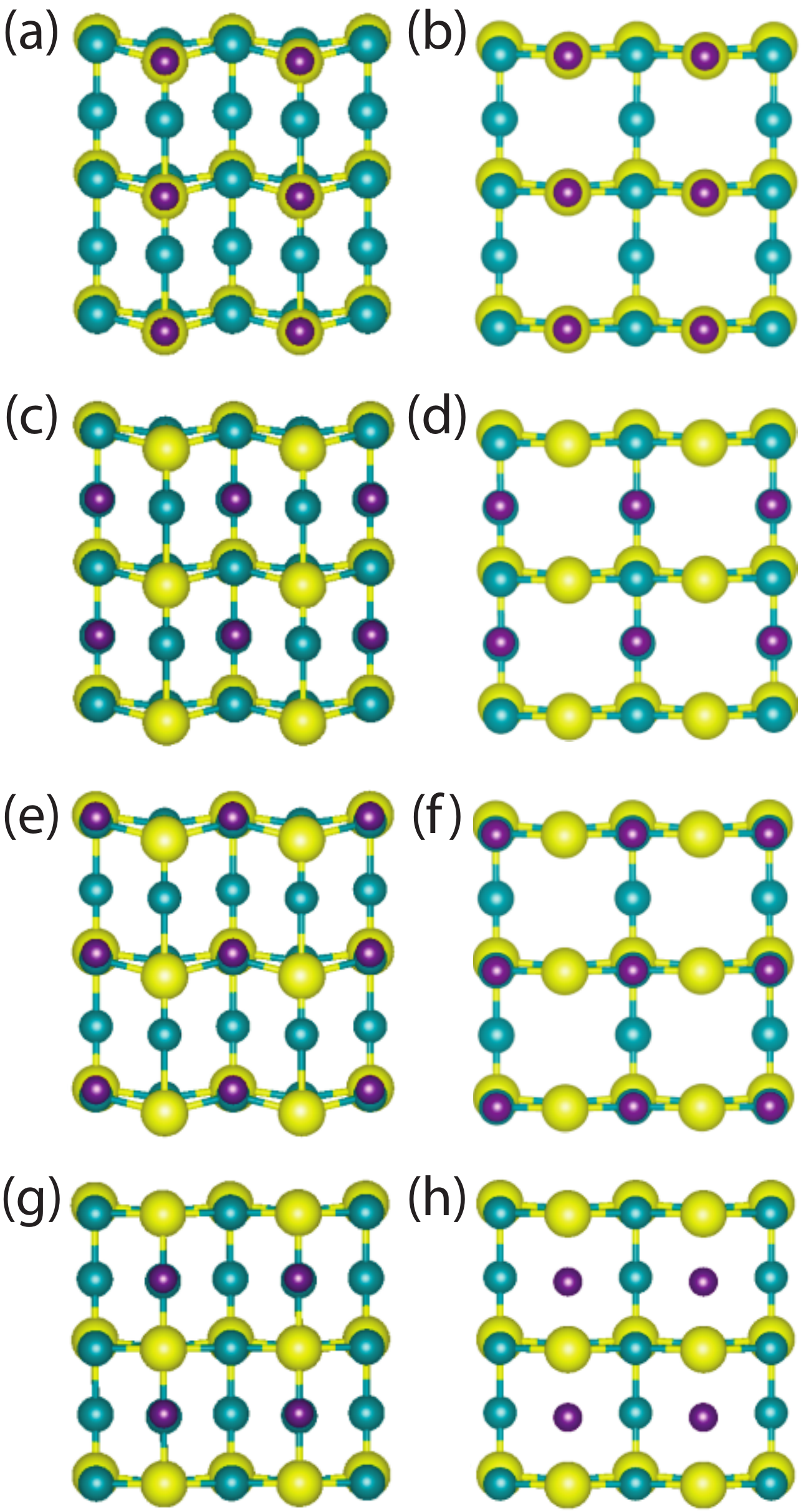}
\end{center}
 \caption{The positions of Se atoms on top of the 2TiO$_x$ surface without (a),(c),(e),(g) and with (b),(d),(f),(h) oxygen vacancies. Yellow, blue, and violet colors indicate Ti, O, and Se atoms, respectively.}
\label{fig:1}
\end{figure}

Oxygen vacancies play an important role, so we study the band structure of the FeSe/\STO system with and without O vacancies in the double TiO layer. To show the general effect of vacancies and to make the calculations manageable, we consider the case of one vacancy per two unit cells. This corresponds to 50\% of vacancies and represents a limiting case, i.e. the result for the fractional number of vacancies $x$ would lie somewhere between the results for $x=0$ and $x=0.5$. Thus, formally, we are dealing with the $\mathrm{FeSe}/\mathrm{Se}/(2\mathrm{TiO}_x)\mathrm{SrTiO}_3$ heterostructure. Since from the experimental study \cite{Zhao2018} the exact positions of Se atoms are not known, we considered several cases shown in Fig.~\ref{fig:1}.
That is, Se is located above Ti atoms [Fig.~\ref{fig:1}(a),(b)], above lower [Fig.~\ref{fig:1}(c),(d)] and upper O [Fig.~\ref{fig:1}(e),(f),(g)] atoms of the double TiO$_x$ layer, and above the oxygen vacancy [Fig.~\ref{fig:1}(h)]. In the case of Se location below the lowest Se of FeSe monolayer, additional configurations was considered with shifted FeSe monolayer in the $xy$ plane so that the additional Se atoms appears right below Fe atoms.

\begin{figure}
 \includegraphics[width=0.95\columnwidth]{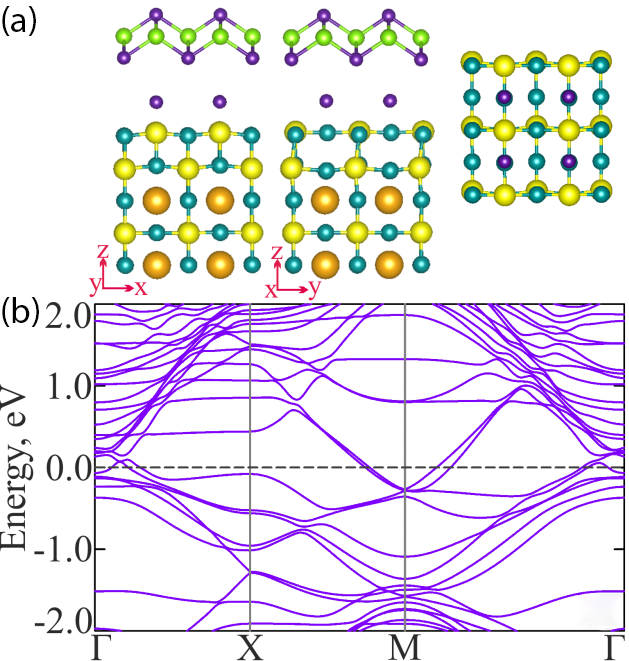}\\\\
 \includegraphics[width=0.95\columnwidth]{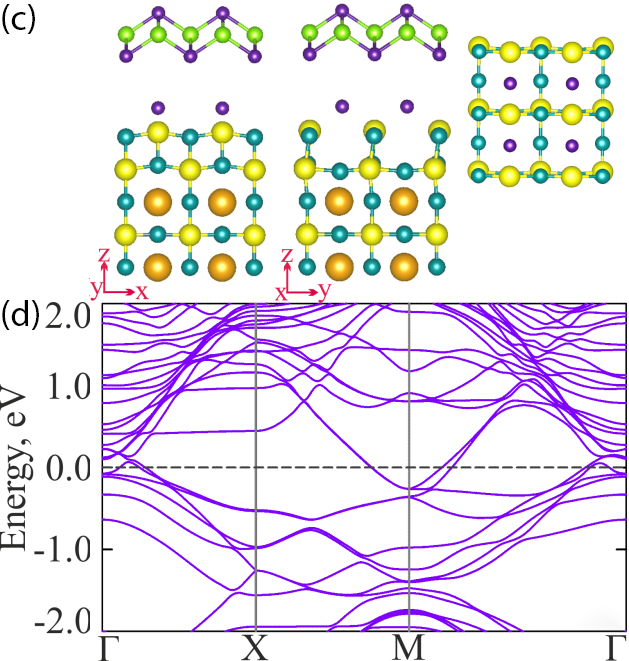}
 \caption{Atomic structures (a),(c) and the corresponding band structures (b),(d) of the energetically favorable configurations of the $\mathrm{FeSe}/\mathrm{Se}/(2\mathrm{TiO}_x)\mathrm{SrTiO}_3$ heterostructure. Fermi level is set to zero. In (a) and (c), yellow, blue, orange, green, and violet colors indicate Ti, O, Sr, Fe, and Se atoms, respectively. For clarity, the substrate along the $z$ axis is only partially shown and along the $x$ and $y$ axes only two unit cells are shown.}
\label{fig:2}
\end{figure}

During the structure optimization, the additional Se atom moved to the position above the upper oxygen atoms of the double TiO layer, Fig.~\ref{fig:1}(g), or above the oxygen vacancies, Fig.~\ref{fig:1}(h). In both cases, location of the additional Se atom under the Fe atoms is energetically favorable. The resulting equilibrium configurations of $\mathrm{FeSe}/\mathrm{Se}/(2\mathrm{TiO}_x)\mathrm{SrTiO}_3$ heterostructure without and with oxygen vacancies are shown in Fig.~\ref{fig:2}(a) and (c), respectively. Corresponding band structures are shown in Fig.~\ref{fig:2}(b) and (d). In the structure without vacancies, the additional Se is above the surface of (2TiO$_x$)\STO by $1.70~\AA$ and below the FeSe monolayer by $2.35~\AA$.
The charge transfer to the FeSe monolayer is $-0.032$ electrons per FeSe formula unit ($\bar{e}$/FeSe). The resulting charge at the additional Se is positive ($+0.058$~$\bar{e}$/Se). As seen from the band structure in Fig.~\ref{fig:2}(b), the additional band crosses the Fermi level around the $\Gamma$ point. It results in the additional Fermi surface electron pocket in the center of the BZ. On the other hand, bands forming the hole pockets are still present and at the $\Gamma$ point they are at 0.23~eV above the $E_F$. It is higher by 0.03~eV than in the reference system without additional Se layer, compare to Fig.~\ref{fig:ref}.

Presence of the oxygen vacancies [50\% in the calculation, see Fig.~\ref{fig:1}(h)] in the upper layer of TiO$_x$ results in the shift of the additional Se towards the substrate layer by $0.22~\AA$. Distance between the FeSe monolayer and the additional layer of Se becomes $3.15~\AA$. Charge transfer to the FeSe monolayer is $-0.045$~$\bar{e}$/FeSe. It is only by $0.013$~$\bar{e}$ more than in the system without vacancies. The charge at the additional Se is $-0.265$~$\bar{e}$/Se. Therefore, the most part of the charge is localized in the additional Se layer. The band structure, shown in Fig.~\ref{fig:2}(d), displays the presence of both hole and electron Fermi surface pockets around the $\Gamma$ point similar to the case without vacancies shown in panel~(b).

From the study of the two energetically favorable cases, we see that the small increase of the electron charge in the FeSe monolayer due to the additional Se layer does not lead to the sinking of the hole bands below the Fermi level at the $\Gamma$ point. On the contrary, new electron pockets formed by the bands with upward dispersion below the Fermi level appear. In general, this is in a disagreement with the ARPES data \cite{FeSeARPES,TanFeSeARPES}. We see that including more details of the FeSe/\STO interface haven't solved the mystery of the hole pocket absence in photoemission. In particular, both oxygen vacancies and additional Se layer certainly affect the band structure near the $\Gamma$ point, but this effect has nothing to do with the simple sinking of hole bands below the Fermi level. Therefore, we have to assume that some other mechanism affects the electronic structure. Natural choice is the effect of strong electronic correlations mentioned in the Introduction. Proper treatment of them requires a separate study. Some success in this direction was achieved within the LDA+DMFT (local density approximation+dynamical mean field theory) studies \cite{Nekrasov2017,Nekrasov2018}

%


\section{Conclusions}
\label{sec:conclusion}

We have studied the effect of the additional Se layer on the band structure of the FeSe/\STO heterostructure. The additional layer is situated between the FeSe monolayer and the double TiO$_x$ layer on top of \STO. For $x=0$ and $x=0.5$, the band structure near the $\Gamma$ point results in the hole Fermi surface pockets and, additionally, a new electron pocket. Therefore, the presence of the additional Se layer does not leads to the sinking of hole bands below the Fermi level. In the presence of the oxygen vacancies in the double TiO layer, Se atoms plays a role of the missing oxygen, i.e. it localizes the charge and prevents its transfer to the FeSe monolayer. We conclude that the strong electronic correlations have to be taken into account to peoperly describe the low energy physics of FeSe monolayers.

\begin{acknowledgements}
The authors would like to thank Information Technology Center, Novosibirsk State University, Institute of Computational Modelling of SB RAS, Krasnoyarsk for providing the access to supercomputer facilities, and Irkutsk Supercomputer Center of SB RAS for providing the access to HPC-cluster ``Akademik V.M. Matrosov'' (Irkutsk Supercomputer Center of SB RAS, Irkutsk: ISDCT SB RAS; \url{http://hpc.icc.ru}, accessed 13.05.2019).
\end{acknowledgements}

%
%

\bibliographystyle{spphys}       
\bibliography{mmkbibl6}   

\begin{thebibliography}{10}
\providecommand{\url}[1]{{#1}}
\providecommand{\urlprefix}{URL }
\expandafter\ifx\csname urlstyle\endcsname\relax
  \providecommand{\doi}[1]{DOI \discretionary{}{}{}#1}\else
  \providecommand{\doi}{DOI \discretionary{}{}{}\begingroup
  \urlstyle{rm}\Url}\fi

\bibitem{y_kamihara_08}
Y.~Kamihara, T.~Watanabe, M.~Hirano, H.~Hosono, Journal of the American
  Chemical Society \textbf{130}(11), 3296 (2008).
\newblock \doi{10.1021/ja800073m}.
\newblock \urlprefix\url{http://dx.doi.org/10.1021/ja800073m}

\bibitem{SadovskiiReview2008}
M.V. Sadovskii, Phys. Usp. \textbf{51}(12), 1201 (2008).
\newblock \doi{10.1070/PU2008v051n12ABEH006820}.
\newblock \urlprefix\url{https://ufn.ru/en/articles/2008/12/b/}

\bibitem{IzyumovReview2008}
Y.A. Izyumov, E.Z. Kurmaev, Phys. Usp. \textbf{51}(12), 1261 (2008).
\newblock \doi{10.1070/PU2008v051n12ABEH006733}.
\newblock \urlprefix\url{https://ufn.ru/en/articles/2008/12/d/}

\bibitem{IvanovskiiReview2008}
A.L. Ivanovskii, Phys. Usp. \textbf{51}(12), 1229 (2008).
\newblock \doi{10.1070/PU2008v051n12ABEH006703}.
\newblock \urlprefix\url{https://ufn.ru/en/articles/2008/12/c/}

\bibitem{JohnstonReview}
D.C. Johnston, Advances in Physics \textbf{59}(6), 803 (2010).
\newblock \doi{10.1080/00018732.2010.513480}.
\newblock \urlprefix\url{http://dx.doi.org/10.1080/00018732.2010.513480}

\bibitem{PaglioneReview}
J.~Paglione, R.L. Greene, Nat. Phys. \textbf{6}(9), 645 (2010).
\newblock \urlprefix\url{http://dx.doi.org/10.1038/nphys1759}

\bibitem{LumsdenReview}
M.D. Lumsden, A.D. Christianson, Journal of Physics: Condensed Matter
  \textbf{22}(20), 203203 (2010).
\newblock \urlprefix\url{http://stacks.iop.org/0953-8984/22/i=20/a=203203}

\bibitem{StewartReview}
G.R. Stewart, Rev. Mod. Phys. \textbf{83}, 1589 (2011).
\newblock \doi{10.1103/RevModPhys.83.1589}.
\newblock \urlprefix\url{http://link.aps.org/doi/10.1103/RevModPhys.83.1589}

\bibitem{Inosov2016}
D.S. Inosov, Comptes Rendus Physique \textbf{17}(1-2), 60  (2016).
\newblock \doi{10.1016/j.crhy.2015.03.001}.
\newblock
  \urlprefix\url{http://www.sciencedirect.com/science/article/pii/S1631070515000523}

\bibitem{HirschfeldKorshunov2011}
P.J. Hirschfeld, M.M. Korshunov, I.I. Mazin, Reports on Progress in Physics
  \textbf{74}(12), 124508 (2011).
\newblock \urlprefix\url{http://stacks.iop.org/0034-4885/74/i=12/a=124508}

\bibitem{BerkSchrieffer}
N.F. Berk, J.R. Schrieffer, Phys. Rev. Lett. \textbf{17}, 433 (1966).
\newblock \doi{10.1103/PhysRevLett.17.433}.
\newblock \urlprefix\url{http://link.aps.org/doi/10.1103/PhysRevLett.17.433}

\bibitem{ChubukovReview2012}
A.~Chubukov, Annual Review of Condensed Matter Physics \textbf{3}(1), 57
  (2012).
\newblock \doi{10.1146/annurev-conmatphys-020911-125055}.
\newblock
  \urlprefix\url{http://dx.doi.org/10.1146/annurev-conmatphys-020911-125055}

\bibitem{Korshunov2014eng}
M.M. Korshunov, Physics-Uspekhi \textbf{57}(8), 813 (2014).
\newblock \doi{10.3367/UFNe.0184.201408h.0882}.
\newblock \urlprefix\url{http://stacks.iop.org/1063-7869/57/i=8/a=813}

\bibitem{KorshunovEreminResonance2008}
M.M. Korshunov, I.~Eremin, Phys. Rev. B \textbf{78}, 140509 (2008).
\newblock \doi{10.1103/PhysRevB.78.140509}.
\newblock \urlprefix\url{http://link.aps.org/doi/10.1103/PhysRevB.78.140509}

\bibitem{Maier2008}
T.A. Maier, D.J. Scalapino, Phys. Rev. B \textbf{78}, 020514 (2008).
\newblock \doi{10.1103/PhysRevB.78.020514}.
\newblock \urlprefix\url{http://link.aps.org/doi/10.1103/PhysRevB.78.020514}

\bibitem{Maier2009}
T.A. Maier, S.~Graser, D.J. Scalapino, P.~Hirschfeld, Phys. Rev. B \textbf{79},
  134520 (2009).
\newblock \doi{10.1103/PhysRevB.79.134520}.
\newblock \urlprefix\url{http://link.aps.org/doi/10.1103/PhysRevB.79.134520}

\bibitem{KorshunovPRB2016}
M.M. Korshunov, V.A. Shestakov, Y.N. Togushova, Phys. Rev. B \textbf{94},
  094517 (2016).
\newblock \doi{10.1103/PhysRevB.94.094517}.
\newblock \urlprefix\url{http://link.aps.org/doi/10.1103/PhysRevB.94.094517}

\bibitem{Korshunov2019}
M.M. Korshunov, Phys. Rev. B \textbf{98}, 104510 (2018).
\newblock \doi{10.1103/PhysRevB.98.104510}.
\newblock \urlprefix\url{https://link.aps.org/doi/10.1103/PhysRevB.98.104510}

\bibitem{Akbari2010}
A.~Akbari, I.~Eremin, P.~Thalmeier, Phys. Rev. B \textbf{81}, 014524 (2010).
\newblock \doi{10.1103/PhysRevB.81.014524}.
\newblock \urlprefix\url{https://link.aps.org/doi/10.1103/PhysRevB.81.014524}

\bibitem{Wang2010}
F.~Wang, H.~Zhai, D.H. Lee, Phys. Rev. B \textbf{81}, 184512 (2010).
\newblock \doi{10.1103/PhysRevB.81.184512}.
\newblock \urlprefix\url{http://link.aps.org/doi/10.1103/PhysRevB.81.184512}

\bibitem{AkbariQPI2010}
A.~Akbari, J.~Knolle, I.~Eremin, R.~Moessner, Phys. Rev. B \textbf{82}, 224506
  (2010).
\newblock \doi{10.1103/PhysRevB.82.224506}.
\newblock \urlprefix\url{https://link.aps.org/doi/10.1103/PhysRevB.82.224506}

\bibitem{Hirschfeld2015}
P.J. Hirschfeld, D.~Altenfeld, I.~Eremin, I.I. Mazin, Phys. Rev. B \textbf{92},
  184513 (2015).
\newblock \doi{10.1103/PhysRevB.92.184513}.
\newblock \urlprefix\url{https://link.aps.org/doi/10.1103/PhysRevB.92.184513}

\bibitem{Mazin2008}
I.I. Mazin, D.J. Singh, M.D. Johannes, M.H. Du, Phys. Rev. Lett. \textbf{101},
  057003 (2008).
\newblock \doi{10.1103/PhysRevLett.101.057003}.
\newblock
  \urlprefix\url{http://link.aps.org/doi/10.1103/PhysRevLett.101.057003}

\bibitem{Chubukov2008}
A.V. Chubukov, D.V. Efremov, I.~Eremin, Phys. Rev. B \textbf{78}, 134512
  (2008).
\newblock \doi{10.1103/PhysRevB.78.134512}.
\newblock \urlprefix\url{http://link.aps.org/doi/10.1103/PhysRevB.78.134512}

\bibitem{Kuroki2008}
K.~Kuroki, S.~Onari, R.~Arita, H.~Usui, Y.~Tanaka, H.~Kontani, H.~Aoki, Phys.
  Rev. Lett. \textbf{101}, 087004 (2008).
\newblock \doi{10.1103/PhysRevLett.101.087004}.
\newblock
  \urlprefix\url{http://link.aps.org/doi/10.1103/PhysRevLett.101.087004}

\bibitem{Graser2009}
S.~Graser, T.~Maier, P.~Hirschfeld, D.~Scalapino, New Journal of Physics
  \textbf{11}(2), 025016 (2009).
\newblock \urlprefix\url{http://stacks.iop.org/1367-2630/11/i=2/a=025016}

\bibitem{MaitiKorshunovPRL2011}
S.~Maiti, M.M. Korshunov, T.A. Maier, P.J. Hirschfeld, A.V. Chubukov, Phys.
  Rev. Lett. \textbf{107}, 147002 (2011).
\newblock \doi{10.1103/PhysRevLett.107.147002}.
\newblock
  \urlprefix\url{http://link.aps.org/doi/10.1103/PhysRevLett.107.147002}

\bibitem{MaitiKorshunovPRB2011}
S.~Maiti, M.M. Korshunov, T.A. Maier, P.J. Hirschfeld, A.V. Chubukov, Phys.
  Rev. B \textbf{84}, 224505 (2011).
\newblock \doi{10.1103/PhysRevB.84.224505}.
\newblock \urlprefix\url{http://link.aps.org/doi/10.1103/PhysRevB.84.224505}

\bibitem{MaitiKorshunov2012}
S.~Maiti, M.M. Korshunov, A.V. Chubukov, Phys. Rev. B \textbf{85}, 014511
  (2012).
\newblock \doi{10.1103/PhysRevB.85.014511}.
\newblock \urlprefix\url{http://link.aps.org/doi/10.1103/PhysRevB.85.014511}

\bibitem{Classen2017}
L.~Classen, R.Q. Xing, M.~Khodas, A.V. Chubukov, Phys. Rev. Lett. \textbf{118},
  037001 (2017).
\newblock \doi{10.1103/PhysRevLett.118.037001}.
\newblock
  \urlprefix\url{https://link.aps.org/doi/10.1103/PhysRevLett.118.037001}

\bibitem{Kontani}
H.~Kontani, S.~Onari, Phys. Rev. Lett. \textbf{104}, 157001 (2010).
\newblock \doi{10.1103/PhysRevLett.104.157001}.
\newblock
  \urlprefix\url{http://link.aps.org/doi/10.1103/PhysRevLett.104.157001}

\bibitem{Onari2012}
S.~Onari, H.~Kontani, Phys. Rev. Lett. \textbf{109}, 137001 (2012).
\newblock \doi{10.1103/PhysRevLett.109.137001}.
\newblock
  \urlprefix\url{https://link.aps.org/doi/10.1103/PhysRevLett.109.137001}

\bibitem{FeSeTc}
W.~Qing-Yan, L.~Zhi, Z.~Wen-Hao, Z.~Zuo-Cheng, Z.~Jin-Song, L.~Wei, D.~Hao,
  O.~Yun-Bo, D.~Peng, C.~Kai, W.~Jing, S.~Can-Li, H.~Ke, J.~Jin-Feng,
  J.~Shuai-Hua, W.~Ya-Yu, W.~Li-Li, C.~Xi, M.~Xu-Cun, X.~Qi-Kun, Chinese
  Physics Letters \textbf{29}(3), 037402 (2012).
\newblock \urlprefix\url{http://stacks.iop.org/0256-307X/29/i=3/a=037402}

\bibitem{Zhang2015}
Z.~Zhang, Y.H. Wang, Q.~Song, C.~Liu, R.~Peng, K.~Moler, D.~Feng, Y.~Wang,
  Science Bulletin \textbf{60}(14), 1301 (2015).
\newblock \doi{https://doi.org/10.1007/s11434-015-0842-8}.
\newblock
  \urlprefix\url{http://www.sciencedirect.com/science/article/pii/S2095927316303711}

\bibitem{GeFeSe100K}
J.F. Ge, Z.L. Liu, C.~Liu, C.L. Gao, D.~Qian, Q.K. Xue, Y.~Liu, J.F. Jia, Nat
  Mater \textbf{14}(3), 285 (2015).
\newblock \urlprefix\url{http://dx.doi.org/10.1038/nmat4153}

\bibitem{ZhaoFeSeLiOHFeSe}
L.~Zhao, A.~Liang, D.~Yuan, Y.~Hu, D.~Liu, J.~Huang, S.~He, B.~Shen, Y.~Xu,
  X.~Liu, L.~Yu, G.~Liu, H.~Zhou, Y.~Huang, X.~Dong, F.~Zhou, K.~Liu, Z.~Lu,
  Z.~Zhao, C.~Chen, Z.~Xu, X.J. Zhou, Nat Commun \textbf{7}, 1 (2016).
\newblock \urlprefix\url{http://dx.doi.org/10.1038/ncomms10608}

\bibitem{Sadovskii2016}
M.V. Sadovskii, Phys. Usp. \textbf{59}(10), 947 (2016).
\newblock \doi{10.3367/UFNe.2016.06.037825}.
\newblock \urlprefix\url{https://ufn.ru/en/articles/2016/10/b/}

\bibitem{Lee2014}
J.J. Lee, F.T. Schmitt, R.G. Moore, S.~Johnston, Y.T. Cui, W.~Li, M.~Yi, Z.K.
  Liu, M.~Hashimoto, Y.~Zhang, D.H. Lu, T.P. Devereaux, D.H. Lee, Z.X. Shen,
  Nature \textbf{515}, 245 (2014).
\newblock \doi{10.1038/nature13894}.
\newblock \urlprefix\url{https://doi.org/10.1038/nature13894}

\bibitem{Coh2015}
S.~Coh, M.L. Cohen, S.G. Louie, New Journal of Physics \textbf{17}(7), 073027
  (2015).
\newblock \doi{10.1088/1367-2630/17/7/073027}.
\newblock \urlprefix\url{https://doi.org/10.1088/1367-2630/17/7/073027}

\bibitem{Gorkov2016}
L.P. Gor'kov, Phys. Rev. B \textbf{93}, 060507 (2016).
\newblock \doi{10.1103/PhysRevB.93.060507}.
\newblock \urlprefix\url{https://link.aps.org/doi/10.1103/PhysRevB.93.060507}

\bibitem{Kulic2017}
M.L. Kuli{\'{c}}, O.V. Dolgov, New Journal of Physics \textbf{19}(1), 013020
  (2017).
\newblock \doi{10.1088/1367-2630/19/1/013020}.
\newblock \urlprefix\url{https://doi.org/10.1088/1367-2630/19/1/013020}

\bibitem{Kang2016}
J.~Kang, R.M. Fernandes, Phys. Rev. Lett. \textbf{117}, 217003 (2016).
\newblock \doi{10.1103/PhysRevLett.117.217003}.
\newblock
  \urlprefix\url{https://link.aps.org/doi/10.1103/PhysRevLett.117.217003}

\bibitem{FernandesReview2017}
R.M. Fernandes, A.V. Chubukov, Reports on Progress in Physics \textbf{80}(1),
  014503 (2017).
\newblock \urlprefix\url{http://stacks.iop.org/0034-4885/80/i=1/a=014503}

\bibitem{Kang2018}
J.~Kang, R.M. Fernandes, A.~Chubukov, Phys. Rev. Lett. \textbf{120}, 267001
  (2018).
\newblock \doi{10.1103/PhysRevLett.120.267001}.
\newblock
  \urlprefix\url{https://link.aps.org/doi/10.1103/PhysRevLett.120.267001}

\bibitem{Yamakawa2017}
Y.~Yamakawa, H.~Kontani, Phys. Rev. B \textbf{96}, 045130 (2017).
\newblock \doi{10.1103/PhysRevB.96.045130}.
\newblock \urlprefix\url{https://link.aps.org/doi/10.1103/PhysRevB.96.045130}

\bibitem{Kreisel2017}
A.~Kreisel, B.M. Andersen, P.O. Sprau, A.~Kostin, J.C.S. Davis, P.J.
  Hirschfeld, Phys. Rev. B \textbf{95}, 174504 (2017).
\newblock \doi{10.1103/PhysRevB.95.174504}.
\newblock \urlprefix\url{https://link.aps.org/doi/10.1103/PhysRevB.95.174504}

\bibitem{Chen2015}
X.~Chen, S.~Maiti, A.~Linscheid, P.J. Hirschfeld, Phys. Rev. B \textbf{92},
  224514 (2015).
\newblock \doi{10.1103/PhysRevB.92.224514}.
\newblock \urlprefix\url{https://link.aps.org/doi/10.1103/PhysRevB.92.224514}

\bibitem{Du2017}
Z.~Du, X.~Yang, D.~Altenfeld, Q.~Gu, H.~Yang, I.~Eremin, P.~Hirschfeld, I.I.
  Mazin, H.~Lin, X.~Zhu, H.H. Wen, Nature Physics \textbf{14}, 134 (2017).
\newblock \doi{10.1038/nphys4299}.
\newblock \urlprefix\url{https://doi.org/10.1038/nphys4299}

\bibitem{Liu2019}
C.~Liu, Z.~Wang, Y.~Gao, X.~Liu, Y.~Liu, Q.H. Wang, J.~Wang, Phys. Rev. Lett.
  \textbf{123}, 036801 (2019).
\newblock \doi{10.1103/PhysRevLett.123.036801}.
\newblock
  \urlprefix\url{https://link.aps.org/doi/10.1103/PhysRevLett.123.036801}

\bibitem{Jandke2019}
J.~Jandke, F.~Yang, P.~Hlobil, T.~Engelhardt, D.~Rau, K.~Zakeri, C.~Gao,
  J.~Schmalian, W.~Wulfhekel, Phys. Rev. B \textbf{100}, 020503 (2019).
\newblock \doi{10.1103/PhysRevB.100.020503}.
\newblock \urlprefix\url{https://link.aps.org/doi/10.1103/PhysRevB.100.020503}

\bibitem{FeSeARPES}
D.~Liu, W.~Zhang, D.~Mou, J.~He, Y.B. Ou, Q.Y. Wang, Z.~Li, L.~Wang, L.~Zhao,
  S.~He, Y.~Peng, X.~Liu, C.~Chen, L.~Yu, G.~Liu, X.~Dong, J.~Zhang, C.~Chen,
  Z.~Xu, J.~Hu, X.~Chen, X.~Ma, Q.~Xue, X.J. Zhou, Nat. Commun. \textbf{3}, 931
  (2012).
\newblock \urlprefix\url{http://dx.doi.org/10.1038/ncomms1946}

\bibitem{TanFeSeARPES}
S.~Tan, Y.~Zhang, M.~Xia, Z.~Ye, F.~Chen, X.~Xie, R.~Peng, D.~Xu, Q.~Fan,
  H.~Xu, J.~Jiang, T.~Zhang, X.~Lai, T.~Xiang, J.~Hu, B.~Xie, D.~Feng, Nat
  Mater \textbf{12}(7), 634 (2013).
\newblock \urlprefix\url{http://dx.doi.org/10.1038/nmat3654}

\bibitem{LiuPRB2012}
K.~Liu, Z.Y. Lu, T.~Xiang, Phys. Rev. B \textbf{85}, 235123 (2012).
\newblock \doi{10.1103/PhysRevB.85.235123}.
\newblock \urlprefix\url{https://link.aps.org/doi/10.1103/PhysRevB.85.235123}

\bibitem{Li2DMat2016}
F.~Li, Q.~Zhang, C.~Tang, C.~Liu, J.~Shi, C.~Nie, G.~Zhou, Z.~Li, W.~Zhang,
  C.L. Song, K.~He, S.~Ji, S.~Zhang, L.~Gu, L.~Wang, X.C. Ma, Q.K. Xue, 2D
  Materials \textbf{3}(2), 024002 (2016).
\newblock \doi{10.1088/2053-1583/3/2/024002}.
\newblock \urlprefix\url{https://doi.org/10.1088/2053-1583/3/2/024002}

\bibitem{Zou2016}
K.~Zou, S.~Mandal, S.D. Albright, R.~Peng, Y.~Pu, D.~Kumah, C.~Lau, G.H. Simon,
  O.E. Dagdeviren, X.~He, I.~Bozovi\'{c}, U.D. Schwarz, E.I. Altman, D.~Feng,
  F.J. Walker, S.~Ismail-Beigi, C.H. Ahn, Phys. Rev. B \textbf{93}, 180506
  (2016).
\newblock \doi{10.1103/PhysRevB.93.180506}.
\newblock \urlprefix\url{https://link.aps.org/doi/10.1103/PhysRevB.93.180506}

\bibitem{Bang2013}
J.~Bang, Z.~Li, Y.Y. Sun, A.~Samanta, Y.Y. Zhang, W.~Zhang, L.~Wang, X.~Chen,
  X.~Ma, Q.K. Xue, S.B. Zhang, Phys. Rev. B \textbf{87}, 220503 (2013).
\newblock \doi{10.1103/PhysRevB.87.220503}.
\newblock \urlprefix\url{https://link.aps.org/doi/10.1103/PhysRevB.87.220503}

\bibitem{Xu2017}
M.~Xu, X.~Song, H.~Wang, Phys. Chem. Chem. Phys. \textbf{19}, 7964 (2017).
\newblock \doi{10.1039/C7CP00173H}.
\newblock \urlprefix\url{http://dx.doi.org/10.1039/C7CP00173H}

\bibitem{Zhao2018}
W.~Zhao, M.~Li, C.Z. Chang, J.~Jiang, L.~Wu, C.~Liu, J.S. Moodera, Y.~Zhu,
  M.H.W. Chan, Science Advances \textbf{4}(3) (2018).
\newblock \doi{10.1126/sciadv.aao2682}.
\newblock \urlprefix\url{https://advances.sciencemag.org/content/4/3/eaao2682}

\bibitem{p_hohenberg_64}
P.~Hohenberg, W.~Kohn, Phys. Rev. \textbf{136}(3B), B864 (1964).
\newblock \doi{10.1103/PhysRev.136.B864}

\bibitem{wkohn65}
W.~{Kohn}, L.J. {Sham}, Phys. Rev. \textbf{140}, 1133 (1965).
\newblock \doi{10.1103/PhysRev.140.A1133}

\bibitem{Ozaki2004}
T.~Ozaki, H.~Kino, Phys. Rev. B \textbf{69}, 195113 (2004).
\newblock \doi{10.1103/PhysRevB.69.195113}.
\newblock \urlprefix\url{https://link.aps.org/doi/10.1103/PhysRevB.69.195113}

\bibitem{Bachelet1982}
G.B. Bachelet, D.R. Hamann, M.~Schl\"uter, Phys. Rev. B \textbf{26}, 4199
  (1982).
\newblock \doi{10.1103/PhysRevB.26.4199}.
\newblock \urlprefix\url{https://link.aps.org/doi/10.1103/PhysRevB.26.4199}

\bibitem{Kleinman1982}
L.~Kleinman, D.M. Bylander, Phys. Rev. Lett. \textbf{48}, 1425 (1982).
\newblock \doi{10.1103/PhysRevLett.48.1425}.
\newblock \urlprefix\url{https://link.aps.org/doi/10.1103/PhysRevLett.48.1425}

\bibitem{Blochl1990}
P.E. Bl\"ochl, Phys. Rev. B \textbf{41}, 5414 (1990).
\newblock \doi{10.1103/PhysRevB.41.5414}.
\newblock \urlprefix\url{https://link.aps.org/doi/10.1103/PhysRevB.41.5414}

\bibitem{Troullier1991}
N.~Troullier, J.L. Martins, Phys. Rev. B \textbf{43}, 1993 (1991).
\newblock \doi{10.1103/PhysRevB.43.1993}.
\newblock \urlprefix\url{https://link.aps.org/doi/10.1103/PhysRevB.43.1993}

\bibitem{Morrison1993}
I.~Morrison, D.M. Bylander, L.~Kleinman, Phys. Rev. B \textbf{47}, 6728 (1993).
\newblock \doi{10.1103/PhysRevB.47.6728}.
\newblock \urlprefix\url{https://link.aps.org/doi/10.1103/PhysRevB.47.6728}

\bibitem{jperdew96}
J.P. {Perdew}, K.~{Burke}, M.~{Ernzerhof}, Phys. Rev. Lett. \textbf{77}, 3865
  (1996).
\newblock \doi{10.1103/PhysRevLett.77.3865}

\bibitem{Momma2011}
K.~Momma, F.~Izumi, Journal of Applied Crystallography \textbf{44}(6), 1272
  (2011).
\newblock \doi{10.1107/S0021889811038970}.
\newblock \urlprefix\url{https://doi.org/10.1107/S0021889811038970}

\bibitem{Nekrasov2017}
I.A. Nekrasov, N.S. Pavlov, M.V. Sadovskii, JETP Letters \textbf{105}(6), 370
  (2017).
\newblock \doi{10.1134/S0021364017060029}.
\newblock \urlprefix\url{https://doi.org/10.1134/S0021364017060029}

\bibitem{Nekrasov2018}
I.A. Nekrasov, N.S. Pavlov, M.V. Sadovskii, Journal of Experimental and
  Theoretical Physics \textbf{126}(4), 485 (2018).
\newblock \doi{10.1134/S1063776118040106}.
\newblock \urlprefix\url{https://doi.org/10.1134/S1063776118040106}

\end{thebibliography}


\end{document}